\documentclass[epsfig,12pt]{article}
\usepackage{epsfig}
\usepackage{graphicx}
\usepackage{rotating}
\usepackage{latexsym}
\usepackage{amsmath}
\usepackage{amssymb}
\usepackage{relsize}
\usepackage{geometry}
\usepackage{color}
\usepackage{bm}
\usepackage{slashed}
\usepackage[normalem]{ulem}

\def\beq{\begin{equation}}
\def\eeq{\end{equation}}
\def\beqn{\begin{eqnarray}}
\def\eeqn{\end{eqnarray}}
\def\beqn{\begin{eqnarray}}
\def\eeqn{\end{eqnarray}}

\def\ba{\beq\new\begin{array}{c}}
\def\ea{\end{array}\eeq}

\newcommand{\cell}{{\mathcal L}}

\newcommand{\p}{\partial}
\newcommand{\wt}{\widetilde}
\newcommand{\ov}{\overline}
\newcommand{\mc}[1]{\mathcal{#1}}
\newcommand{\md}{\mathcal{D}}
\newcommand{\ml}{\mathcal{L}}

\newcommand{\lgr}{\left\lgroup}
\newcommand{\rgr}{\right\rgroup}

\newcommand{\ii}{\hat\imath}
\newcommand{\jj}{\hat\jmath}
\newcommand{\kk}{\hat k}

\newcommand{\cR}{\mathcal{R}}
\newcommand{\cC}{\mathcal{C}}
\newcommand{\cH}{\mathcal{H}}
\newcommand{\cO}{\mathcal{O}}

\begin{document}

\begin{titlepage}
\ \ 
\vskip 2cm
\begin{center}
{  \Large \bf  Quaternionic Wavefunction }
\end{center}
\vskip 0.5cm

\begin{center}

 {\large
 \bf   Pavel A.~Bolokhov
 }
\end {center}

\begin{center}
{\it Theoretical Physics Department, St.Petersburg State University, Ulyanovskaya~1, 
	Peterhof, St.Petersburg, 198504, Russia}
\end{center}

\begin{center}
{\large\bf Abstract}
\end{center}

\hspace{0.3cm}
	We argue that quaternions form a natural language for the description of
	quantum-mechanical wavefunctions with spin.
	We use the quaternionic spinor formalism which is in one-to-one correspondence with the usual
	spinor language.
	No unphysical degrees of freedom are admitted, in contrast to the majority of literature on quaternions.
	In this paper we first build a Dirac Lagrangian in the quaternionic form,
	derive the Dirac equation and take the non-relativistic limit to find the Schr\"odinger's equation.
	We show that the quaternionic formalism is a natural choice to start with,
	while in the transition to the non-interacting non-relativistic limit the quaternionic description 
	effectively reduces to the regular complex wavefunction language.
	We provide an easy to use grammar for switching between the ordinary spinor language
	and the description in terms of quaternions.
	As an illustration of the broader range of the formalism,
	we also derive the Maxwell's equation from the quaternionic Lagrangian of Quantum Electrodynamics.
	In order to derive the equations of motion, we develop the variational calculus
	appropriate for this formalism.
\vspace{2cm}

\end{titlepage}

\section{Introduction}
\setcounter{equation}{0}

	The list of literature on the r\^ole of quaternions in physics is so vast that it is hardly
	possible to enlist it \cite{Adler:1988hb}--\cite{Chanyal:2012rj}.
	The most of the literature touches base on the use of quaternions in three-dimensional
	rotations and Lorentz transformations.
	Other works include their applications in Electrodynamics and Quantum Mechanics.
	The attractive feature of quaternions is that whenever a solution is possible in the quaternionic form,
	it is much more compact than that in the customary form of Lorentz vectors and tensors. 

	For the majority of theorists, however, quaternions are seen as a somewhat exotic subject,
	which neither has proven to be exceedingly effective, nor has lead
	to any new insights or new formalisms.

	The other drawback, as it is perceived, is a somewhat ``strange'' mathematical language,
        often accompanied by strange results following from it \cite{DeLeo:1998yi}.
	While from learning Quantum Mechanics and Gauge Theories we are used to non-commutative operators,
	and non-commutative objects in general, when we see an operator of multiplication ``from the right'' $ |\, \ii $
        (an example of the so-called ``barred'' operators) it immediately induces a certain degree of skepsis.
	That is, mathematically this formalism may be interesting, but physically this seems to be
	driven away from reality and therefore deemed unnecessary.

	In the literature, quaternionic wavefunction is usually introduced \emph{ad hoc}.
	The wavefunction is named quaternionic just as an attempt of a generalization of Quantum Mechanics \cite{Adler:1988hb}, \cite{Edmonds:1972vd},
	Quantum Electrodynamics \cite{DeLeo:2000ik}, Gravity \cite{Rawat:2011ch} and so on.
	Even though such a theory may involve a new form of analysis, new operator formalisms,
	and other attractive mathematical features, it is not warranted by experiment in any sense. 
	The resulting Dirac or Schr\"odinger equations often include unobserved degrees of freedom.
	This is the consequence of the fact that quaternions are multi-component numbers,
        with a bit too many components than needed for Quantum Mechanics.
        These aspects of using quaternions in theoretical physics are enough to discourage the interest in the
        majority of theorists.

	Recently, a series of works have been published \cite{Furey:2010fm}--\cite{Furey:2015tqa} where a formalism has been developed for constructing
	the quaternionic analogues of spinors, based on the determination of the so-called
	maximally totally isotropic subspaces which project the desired minimal ideals from the
	complex quaternionic algebra $ \cC \otimes \cH $.
	This construction has a direct physical interpretation in terms of chirality of spinors,
	which we actively exploit in this paper.
	This way we are able to ensure that the content of the theory we are writing does not
	include any exotic or non-observed degrees of freedom.

	Our main goal is to cast a bridge from the regular algebraic language of physics (which is, dominantly based
	on complex numbers) to the language of quaternions.
	We argue that quaternions have always been around, and we just neglected to acknowledge them.
	There is no need or necessity for any new degrees of freedom or new physics to arise.
	We would like to present a concise dictionary, so that any theorist could connect to and appreciate
	the quaternionic formalism, which appears to be quite capacious.

	The omnipresence of quaternions is easy to observe.
	We know that Quantum Mechanics is based solely on complex numbers.
	Complex numbers provide a compact and meaningful way of both formulating and solving quantum-mechanical problems.
	With some exceptions, it would be very awkward to split the Schr\"odinger equation into its real and imaginary parts,
	and then to attempt to solve the resulting system of equations.
	Quantum mechanical operators of momentum, angular momentum are \emph{inherently} complex.
	That is, to say, that the wavefunction is complex too.
	
	But as soon as the relativistic effects are included into Quantum Mechanics, it turns out that particles
	have \emph{spin}.
	The way to incorporate spin into the wavefunction is just to turn it into a spinor.
	The spin operator itself is then given by the Pauli matrices.
	This is where quaternions get involved.
	The algebra of Pauli matrices is the same as that of quaternionic units, loosely speaking.
	We argue that there is \emph{no inherent need} to having introduced matrices.
	Quaternionic units are all that is needed, and \emph{implicitly} the wavefunction in Quantum Mechanics is quaternionic.

	So, how does one include spin into the wavefunction by means of quaternions?
	We proceed in a very conservative way, essentially expanding on the development originally presented in \cite{thesis}:
	we start out from the Dirac Lagrangian, carefully taking into account the spin degrees of freedom,
	and then \emph{derive} the Schr\"odinger equation.
	Spin naturally appears as a consequence of the fact that the wavefunction is quaternionic.
	Quaternionic derivations, while a bit unusual for some, are simpler than spinor derivations.
	This way we argue that quaternionic language is natural for Quantum Mechanics.

	A well-known peculiar property of physics is that physical objects are actually described by \emph{complex} quaternions,
	which are also known as complexified quaternions, and not by the regular quaternions.
	This does not change our main argument, however.
	Quaternionic structure allows one pass from the matrix formulation of the theory to the algebraic description.
	In order to work with spinors one normally has to involve Dirac or Pauli matrices, which often conceal
	the underlying algebraic structure of the solution.
	Needless to say, that the equations take a lot more attractive look when such matrices are not involved.

	It is not just for these reasons that we believe quaternions play a fundamental r\^ole in physics.
	There are hints that quaternions are part of the natural language for the entire Standard Model \cite{Furey:2015tqa}.
	We view this work as one of the steps towards the description of the Standard Model in
	such a language.

\vskip 0.8cm
\centerline{*\qquad\qquad\qquad*\qquad\qquad\qquad*}
\vskip 0.6cm

	Let us talk about our notations first, while gradually introducing the subject matter.
	The reader eager to see the physical results may choose to skip to the next section, returning here
	to clarify the notations when necessary.
	Here we overview the known facts which are easy to pick up and use,
	while their proof can be found in the literature \cite{DeLeo:1997yj}.
	We denote the quaternionic units as
\begin{equation}
	\ii^2	~~=~~	\jj^2	~~=~~	\kk^2	~~=~~	-1\,,
\end{equation}
	and we do not distinguish them from the three-dimensional spatial unit vectors.
	That is, any three-dimensional vector is a quaternion
\begin{equation}
	\vec a	~~=~~	a^1\, \ii  ~~+~~  a^2\, \jj  ~~+~~  a^3\, \kk\,.
\end{equation}

	Complex quaternions are defined as
\begin{equation}
\label{cq}
	a	~~=~~	a^0  ~~+~~  a^1 \ii  ~~+~~  a^2 \jj  ~~+~~  a^3 \kk\,,
\end{equation}
	where all components
\begin{align}
	a^0	& ~~=~~	b^0  ~~+~~  i\,b^0\,,
	&
	a^1	& ~~=~~ b^1  ~~+~~  i\,b^1\,,
	&
	\dots,
\end{align}
	are complex numbers.
	Here $ i $ denotes a regular imaginary complex unit, which commutes with $ \ii $, $ \jj $, $ \kk $.

	Let us introduce the conjugation operations.
	We denote the quaternionic conjugation (q.c.) by $ \wt a $,
\begin{equation}
	\wt a	~~=~~	a^0  ~~-~~  a^1 \ii  ~~-~~  a^2 \jj  ~~-~~ a^3 \kk\,.
\end{equation}
	We remember that the quaternionic conjugation switches the order of factors in a product:
\begin{equation}
	\wt{a\, b}  ~~=~~  \wt b\, \wt a\,.
\end{equation}
	Since the components are complex numbers, we also have the complex conjugation (c.c.) $ a^* $
\begin{equation}
	a^*	~~=~~	(a^0)^*  ~~+~~  (a^1)^*\, \ii  ~~+~~  (a^2)^*\, \jj  ~~+~~  (a^3)^*\, \kk\,.
\end{equation}
	It is convenient to introduce the composition of these two conjugations, which we call a \emph{hermitean}
	conjugation (h.c.) (not without a reason),
\begin{equation}
	a^\dag	~~=~~	(a^0)^*  ~~-~~  (a^1)^*\, \ii  ~~-~~  (a^2)^*\, \jj  ~~-~~  (a^3)^*\, \kk\,.
\end{equation}
	By itself it does not give anything new, as it is merely a combination of $ \,\wt{~}\, $ and $ \,*\, $ operations,
	but it is important to have it, as we will see.
	Hermitean conjugation also interchanges the order of terms in a product, obviously
\begin{equation}
	(a\, b)^\dag	~~=~~	b^\dag\, a^\dag\,.
\end{equation}
	It is this plentitude of conjugations that give richness to the quaternionic language, when applied to physics.

	Now we can define a \emph{true} four-dimensional vector
\begin{equation}
\label{true-vector}
	v	~~=~~	v^0  ~+~  i\,(\, v^1\,\ii  ~+~  v^2\,\jj  ~+~ v^3\,\kk \,)
		~~\equiv~~	v^0  ~+~  i\,\vec v
\end{equation}
	in Minkowski space.
	Note that it is precisely the combination $ i\,\vec v $ that gives a true vector.
	Oppositely,
\begin{equation}
\label{pseudo-vector}
	v	~~=~~	i\,v^0  ~+~  \vec v
\end{equation}
	describes a \emph{pseudo}-scalar and a \emph{pseudo}-vector, with respect to parity inversion $ P $.
	Together, the two objects \eqref{true-vector} and \eqref{pseudo-vector} span the entire space of 
	complex quaternions \eqref{cq}.
	In other words, a generic complex quaternion $ a $ can be split into two four-vectors.
	Their time components will represent a true- and a pseudo-scalar, while their three-dimensional
	parts will represent a true and an axial vector, correspondingly.
	Notice, how multiplication by the complex $ i $ turns a true vector into an axial vector, and
	the same for scalars.

	With four-dimensional vectors in Minkowski space, one has to be careful to always keep in mind whether
	a vector is contravariant or covariant.
	Complex conjugation $ * $ turns a contravariant vector $ v $ into a covariant vector $ v^* $.
\begin{equation}
	v^*	~~=~~	v^0  ~-~  i\,\vec v\,.
\end{equation}
	This actually is equivalent to quaternionic conjugation $ v^*  ~=~ \wt v $.

	The derivative operator $ \partial $
\begin{equation}
	\partial	~~=~~	\partial^0  ~+~  i\,\vec\nabla
\end{equation}
	is \emph{by definition} a covariant vector, while obviously
\begin{equation}
	\partial^*	~~=~~	\partial^0  ~-~  i\,\vec\nabla
\end{equation}
	is a contravariant vector.

	To make these identifications more meaningful, let us talk about Lorentz transformations.
	Lorentz transformations are generated by a purely-imaginary quaternionic parameter
\begin{equation}
	\Lambda		~~=~~	\vec\kappa  ~~+~~  i\,\vec\lambda\,.
\end{equation}
	We are not going to explicitly treat it as a vector, so we are not putting a vector sign on this parameter.
	Parameter $ \vec\kappa $ generates three-dimensional rotations, while parameter $ \vec\lambda $
	generates boosts.
	These are ``generators'' in the sense that the actual finite transformations are performed by the exponent
\begin{equation}
	e^\Lambda\,.
\end{equation}
	It is important to be careful here, as $ \vec\kappa $ can be interpreted as a three-dimensional rotation
	only when $ \vec\lambda ~=~ 0 $, and the same for $ \vec\lambda $ --- it can be interpreted as a boost
	only when $ \vec\kappa ~=~ 0 $.
	This is because in general $ \vec\kappa $ and $ \vec\lambda $ do not commute, and therefore
\begin{equation}
	e^{\vec\kappa ~+~ i\,\vec\lambda}	~~\neq~~	e^{\vec\kappa}  ~\cdot~  e^{i\,\vec\lambda}\,,
\end{equation}
	where each individual exponent on the right-hand side {\emph is} treated as a rotation and a boost,
	correspondingly.
	Notice that since $ \Lambda $ is purely imaginary
	(\emph{i.e.} its real part vanishes),
	then $ \wt\Lambda ~=~ -\Lambda $, and therefore
\begin{equation}
	\wt{(e^\Lambda)}		~~=~~	e^{-\Lambda}\,.
\end{equation}
	A contravariant vector $ v $ transforms under $ \Lambda $ as
\begin{equation}
\label{contra}
	v	~~\to~~		e^\Lambda\,v\,e^{\Lambda^\dag}\,.
\end{equation}
	Any covariant vector then should transform the same way that $ v^* $ does:
\begin{equation}
\label{co}
	v^*	~~\to~~		e^{\Lambda^*}\,v^*\,e^{\wt\Lambda}\,.
\end{equation}
	As a special case of these, a three-dimensional vector $ \vec v $ rotates as
\begin{equation}
\label{rotation}
	\vec v	~~\to~~		e^{\vec\kappa}\,\vec v\,e^{-\vec\kappa}\,.
\end{equation}
	This concludes our basic discussion of vectors for now.

	Another way a complex quaternion \eqref{cq} can be split up, is by separating its zeroth component $ a^0 ~\equiv~ \phi $
	and its vector part $ \vec a $.
	The zeroth component $ \phi $ is a complex number whose real and imaginary parts are identified
	as a true and axial scalar fields, correspondingly.
	The remaining vector part is then identified as a field strength:
\begin{equation}
\label{fs}
	\vec F	~~=~~	\vec B  ~+~  i\,\vec E\,.
\end{equation}
	Notice that this agrees with the identifications of \eqref{true-vector} and \eqref{pseudo-vector}
	as polar and axial vectors.
	While both $ \vec B $ and $ \vec E $ are vectors in the three-dimensional sense,
	one does not view the object \eqref{fs} as a four-dimensional vector in any way.
	This is an entirely different split up of a complex quaternion.

	Both $ \phi $ and $ \vec F $ transform the same way under Lorentz transformations:
\begin{equation}
	\qquad\qquad\qquad\qquad
	\rho	~~\to~~	e^\Lambda\,\rho\,e^{\wt\Lambda}\,,
	\qquad\qquad\qquad
	\rho	~=~	\phi\,,~ \vec F\,.
\end{equation}
	For the scalar $ \phi $ this obviously does not do anything, since $ \phi $ is just a complex number:
\[
	e^\Lambda\,\phi\,e^{\wt\Lambda}	~~=~~	\phi\, e^\Lambda\, e^{\wt\Lambda}	~~=~~	\phi\,,
\]
	so it is indeed a scalar.
	While for the field strength, this dictates that
\begin{equation}
	\vec F		~~\to~~		e^\Lambda\,\vec F\,e^{\wt\Lambda}\,.
\end{equation}
	Note that for the case of pure rotations, when $ \Lambda ~=~ \vec\kappa $,
	this agrees with Eq.~\eqref{rotation}, as $ \wt\Lambda ~=~ -\Lambda ~=~ -\vec\kappa $.
	That is, both the electric and magnetic fields rotate as three-vectors.

	Finally, we introduce spinors.
	Here we give a very brief overview of spinors necessary for Section~\ref{section-dirac}, while
	a more detailed discussion is postponed until Appendix~\ref{section-spinors}.
	We begin with stating that the space of complex quaternions $ \cC \otimes \cH $
	can be split in two halves in a yet another way, namely using chirality projectors, $ P_L $ and $ P_R $.
	This construction is based on identifying the projectors corresponding to the
	maximally totally isotropic subspaces of the quaternion algebra \cite{thesis}.
	In the case of $ \cC \otimes \cH $ such spaces are one-dimensional each, and are given
	by the projectors $ P_L $ and $ P_R $.
	A \emph{left-handed} spinor is defined as an arbitrary complex quaternion multiplied by 
	a projector $ P_L $ on the right:
\begin{equation}
	\psi_L	~~=~~	a\,P_L\,.
\end{equation}
	Here $ P_L $ is a complex quaternion
\begin{equation}
	P_L	~~=~~	\frac{1 \,+\, i\kk} 2\,,
\end{equation}
	and we call it a projector because
\[
	P_L^2	~~=~~	P_L\,.
\]
	All accompanying details of these definitions can be found in Appendix~\ref{section-spinors}.
	For an ordinary reader $ \psi_L $ should be precisely viewed as a left-handed
	chiral spinor.
	A convenient basis for $ \psi_L $ is formed by elements $ P_L $ and $ \jj\,P_L $,
	which are geometrically orthogonal to each other:
\begin{equation}
\label{lbasis}
	\psi_L	~~=~~	\xi_L\,P_L  ~~+~~  \chi_L\,\jj\,P_L\,.
\end{equation}
	Complex numbers $ \xi_L $ and $ \chi_L $ are precisely the ``spin-up'' and ``spin-down''
	components of $ \psi_L $ viewed as a Weyl spinor:
\begin{equation}
	\psi_L	~~=~~	\left\lgroup
				\begin{matrix}
					\xi_L \\
					\chi_L
                		\end{matrix}
			\right\rgroup.
\end{equation}

	Left-handed spinors span one half of the complex quaternion space, while the other half
	is spanned by the right-handed spinors:
\begin{equation}
	\psi_R	~~=~~	a\,P_R\,,
\end{equation}
	where
\begin{equation}
	P_R	~~=~~	\frac{1 \,-\, i\kk} 2\,.
\end{equation}
	These two halves are related by complex conjugation $ * $,
	and the projectors are related as
\begin{align}
	P_R	& ~~=~~	P_L^*\,,
	&
	P_L  ~~+~~  P_R	& ~~=~~ 1\,.
\end{align}
	The basis for $ \psi_R $ is similarly given by $ P_R $ and $ \jj\,P_R $, but the
	components are identified slightly differently
\begin{equation}
\label{rbasis}
	\psi_R	~~=~~	-\xi_R\,\jj\,P_R  ~~+~~  \chi_R\,P_R\,.
\end{equation}
	Defined like so,
\begin{equation}
	\psi_R	~~=~~	\left\lgroup
				\begin{matrix}
					\xi_R \\
					\chi_R
                		\end{matrix}
			\right\rgroup
\end{equation}
	is precisely identified as a right-handed Weyl spinor.
	The basis \eqref{lbasis} and especially so \eqref{rbasis} may seem a bit awkward,
	but they are convenient for doing algebra.
	The fact that the projectors are part of the bases allows us to perform
	various manipulations and conjugations on spinors quite effectively.

	Under Lorentz transformations, left spinors by definition transform as
\begin{equation}
	\psi_L	~~\to~~		e^\Lambda\, \psi_L\,,
\end{equation}
	while the right-handed ones transform in the conjugate representation
\begin{equation}
	\psi_R	~~\to~~		e^{\Lambda^*}\, \psi_R\,.
\end{equation}
	It is interesting to observe and compare how spinors and three-dimensional vectors
	transform under three dimensional rotations.
	Consider a rotation around axis $ \hat a $ ($ \hat a{}^2 ~=~ -1 $) through angle $ \alpha $.
	Vector $ \vec v $ rotates as
\begin{equation}
	\vec v	~~\to~~		e^{\alpha\, \hat a/2}\, \vec v\, e^{-\alpha\, \hat a/2}\,,
\end{equation}
	while spinors transform as
\begin{equation}
	\psi_{L,R}	~~\to~~	e^{\alpha\, \hat a/2}\, \psi_{L,R}\,.
\end{equation}
	In particular, if we perform a rotation through $ 2\pi $, then
	$ e^{\pm 2\pi\,\hat a/2} ~=~ -1 $, and a vector is unchanged,
	while a spinor changes its sign, as it should be.
	Of course, this is because of the factor of $ 1/2 $ in the exponent (the famous ``Rodrigues' two''), which
	is just the reflection of the fact that SU(2) is a double cover of
	the rotation group SO(3).

	The action of discrete symmetries on spinors is presented in Appendix~\ref{section-discrete}.

\section{Dirac equation}
\label{section-dirac}
\setcounter{equation}{0}

	Our goal in this section is to construct the Lagrangian for the electron in electromagnetic field,
	in quaternionic form, and derive the equation of motion --- the Dirac equation.

	Let us begin with a massless particle, in the absence of the electromagnetic field.
	The appropriate Lagrangian was given in \cite{thesis},
\begin{equation}
\label{L-massless}
	\mc L_\text{massless}
		~~=~~	\psi_L^\dag\, i\partial\, \psi_L  ~~+~~  \psi_R^\dag\, i\,\partial^*\, \psi_R
			~~+~~  \text{c.c.}
\end{equation}
	Note that Lorentz invariance is manifest here, because
\begin{align}
	\psi_L^\dag	& ~~\to~~	\psi_L^\dag\, e^{\Lambda^\dag}  ~~=~~  \psi_L^\dag\, e^{-\Lambda^*}\,,
	&
	\partial	& ~~\to~~	e^{\Lambda^*}\, \partial\, e^{-\Lambda}\,,
	&
	\psi_L		& ~~\to~~	e^\Lambda\, \psi_L\,,
\end{align}
	and
\begin{align}
	\psi_R^\dag	& ~~\to~~	\psi_R^\dag\, e^{\wt \Lambda}  ~~=~~  \psi_R^\dag\, e^{-\Lambda}\,,
	&
	\partial^*	& ~~\to~~	e^\Lambda\, \partial^*\, e^{-\Lambda^*}\,,
	&
	\psi_R		& ~~\to~~	e^{\Lambda^*}\, \psi_R\,.
\end{align}

	We need to make an important remark about conjugating products of spinors,
	due to the fact that spinors are Grassmann variables.
	By definition, complex conjugation of two Grassmann variable interchanges their order,
\begin{equation}
	(\zeta\, \eta)^*	~~=~~	\eta^*\, \zeta^*\,,\qquad\qquad \text{for Grassmann numbers.}
\end{equation}
	If we take two complex quaternions $ \xi $ and $ \chi $, which are fermions at the same time,
	complex conjugation cannot change their order, because of their quaternionic content.
	In that case, the order is preserved, but an extra minus sign appears,
\begin{equation}
\label{ferm-cc}
	(\xi\, \chi)^*		~~=~~	-\, \xi^*\, \chi^*\,,\qquad \text{for fermionic complex quaternions.}
\end{equation}
	If we take a quaternionic conjugate $ \wt{\xi\, \chi} $, on the other hand, 
	the conjugation will attempt to change their order precisely because of the quaternionic content.
	Note that the quaternionic algebra \emph{requires} us to interchange the factors, or
	the result will simply be incorrect.
	But now because the spinors are fermions, and we are \emph{not} performing a
	complex conjugation, we get an extra minus sign
\begin{equation}
\label{ferm-qc}
	\wt{\xi\, \chi}		~~=~~	-\, \wt\chi\, \wt\xi\,\qquad \text{for quaternionic spinors.}
\end{equation}
	The only kind of conjugation which does not produce a negative sign is hermitean
	conjugation --- this combination changes the order of the spinors in agreement with
	both complex and quaternionic conjugations,
\begin{equation}
\label{ferm-hc}
	(\xi\, \chi)^\dag	~~=~~	\chi^\dag\, \xi^\dag\,.
\end{equation}

	Lagrangian \eqref{L-massless} is manifestly chiral, in that it consists of two separate
	terms for the left-handed and right-handed fermions.
	This might be considered as a disadvantage compared to the usual matrix-based representation,
	where both the mass term and the kinetic term are united.
	In quaternionic language, we keep the chiralities in Eq.~\eqref{L-massless} separate here.
	As we will be dealing with the quantum-mechanics limit, explicit chiral structure is more advantageous
	in this study.
	Furthermore, the Standard Model, the way to which this work is meant to point, is chiral in its nature.
	It is, however, possible to unite the terms in Eq.~\eqref{L-massless}.
	To do that, we need to appeal to the so-called $ m_c $-action \cite{thesis} that the derivatives
	in \eqref{L-massless} implement.
	In plain terms, the kinetic term involves the projectors acting on the right,
\begin{equation}
\label{L-united}
	\mc L_\text{massless}	~~=~~	\psi_D^\dag\, i\!\lgr \p \,|\, P_L ~+~ \p^* \,|\, P_R \rgr\! \psi_D  ~~+~~  \text{c.c.}
\end{equation}
	Expressing the kinetic term in this form will be useful for promoting the theory
	to the quantum stage and developing the Feynman rules.
	This will be the subject of future work.

	Let us discuss gauge transformations now, as we need the electron to interact with the electromagnetic field.
	Gauge transformations just rotate the overall complex phase of a spinor, and so they are defined
	similarly both for right- and left-handed spinors,
\begin{align}
	\psi_L		& ~~\to~~		e^{i\varphi}\,\psi_L\,,
	&
	\psi_R		& ~~\to~~		e^{i\varphi}\,\psi_R\,.
\end{align}
	Notice that the quaternionic conjugates also transform the same way,
\begin{align}
	\wt\psi{}_L		& ~~\to~~		\wt\psi{}_L\,e^{i\varphi}\,,
	&
	\wt\psi{}_R		& ~~\to~~		\wt\psi{}_R\,e^{i\varphi}\,.
\end{align}
	Although $ e^{i\varphi} $ certainly commutes with $ \wt\psi{}_{L,R} $,
	here for convenience we wrote it on the right of the latter.
	Both the complex conjugates and hermitean conjugates will have the opposite charge,
\begin{align}
\notag
	\psi_L^*		& ~~\to~~		\psi_L^*\,e^{-i\varphi}\,,
	&
	\psi_R^*		& ~~\to~~		\psi_R^*\,e^{-i\varphi}\,,
	\\[4mm]
	\psi_L^\dag		& ~~\to~~		\psi_L^\dag\,e^{-i\varphi}\,,
	&
	\psi_R^\dag		& ~~\to~~		\psi_R^\dag\,e^{-i\varphi}\,.
\end{align}

	In order to make this transformation local, we define the long derivative,
\begin{equation}
	\md		~~=~~	\partial  ~~-~~  i\,A^*\,.
\end{equation}
	The reason that we have to put $ A^* $ here instead of just $ A $ is because
	$ \partial $ and $ \md $ are covariant vectors.
	This is just the reflection of the fact that $ A^\mu $ enters the long derivative
	with the lower index $ \mu $:
\begin{equation}
	\md_\mu		~~=~~	\partial_\mu  ~~-~~  i\,A_\mu\,.
\end{equation}
	This long derivative then transforms as
\begin{equation}
	\md		~~\to~~		e^{i\varphi}\, \md\, e^{-i\varphi}\,,
\end{equation}
	meaning that, as usual,
\begin{equation}
	A_\mu		~~\to~~		A_\mu  ~~+~~  \partial_\mu\,\varphi\,.
\end{equation}
	This allows $ \md $ to act on $ \psi_L $, so that $ \md\,\psi_L $ is again in the fundamental representation
	of U(1).
	Now, although $ \psi_R $ has the same charge as $ \psi_L $, we cannot act on it with the same derivative, because
	the product $ \md\,\psi_R $ will not transform under the Lorentz transformations properly.
	Remind, that the right-handed and left-handed spinor spaces are in fact related by complex conjugation.
	This was the reason that we wrote $ \partial^* $ in the Lagrangian in Eq.~\eqref{L-massless}.
	One would think that by analogy we should act on $ \psi_R $ with $ \md^* $ --- but that would also be
	a mistake because it would imply that $ \psi_R $ has the opposite charge.
	In reality, it is the quaternionic conjugate $ \wt\md $ that should be put into the Lagrangian.
	This conjugation does not change the sign of the electric charge.
	Overall, the Lagrangian now looks as,
\begin{equation}
\label{L-gauge}
	\mc L_\text{massless}
		~~=~~	\psi_L^\dag\, i\md\, \psi_L  ~~+~~  \psi_R^\dag\, i\,\wt\md\, \psi_R
			~~+~~  \text{c.c.}
\end{equation}
	Before proceeding, let us emphasize the remarkable feature of the Lagrangians \eqref{L-massless} and \eqref{L-gauge},
	which we could have done earlier: the absence of $ \gamma $-matrices (or $ \sigma $-matrices for that matter).
	The only residue of the matrix structure of the Dirac's Lagrangian is residing in the fact that the Lagrangians
	have two chiral terms, instead of just one.
	We will return to this below.

	Now we add the mass term.
	Since this has to be the Dirac mass, it has to flip chirality.
	The only form that correctly reproduces the mass term has the form $ m\, \jj\, \psi_L^\dag\, \psi_R $,
\begin{equation}
	m\, \jj\, \psi_L^\dag\, \psi_R  ~~+~~  \text{q.c.}  ~~+~~  \text{c.c}  ~~=~~
	m \lgr \jj\, \psi_L^\dag\, \psi_R  ~+~  \wt\psi{}_R\, \psi_L^*\, \jj \rgr
	~~+~~  \text{c.c.}
\end{equation}
	Here the signs q.c. and c.c. imply adding the appropriate conjugate of everything
	that resides to the left of the respective sign.
	We will discuss the occurrence of $ \jj $ in this expression in detail in Appendix~\ref{section-spinors}.
	This form of the mass term seems awkward, and we will be able to get rid of it soon, after we discover its meaning.
	For now we just note that without it the expression would vanish, \emph{i.e.} $ \psi_L^\dag\, \psi_R $
	has no real part.
	Let us also note that this factor of $ \jj $ can be moved to the right at our convenience:
\begin{equation}
\label{m-term}
	m\, \psi_L^\dag\, \psi_R\, \jj  ~~+~~  \text{q.c.}  ~~+~~  \text{c.c.}
\end{equation}
	This follows from the general cyclic property of quaternion product
	under the ``$ \text{q.c.} $'' sign, analogous to the cyclicity of trace of matrices\footnote{
		In fact, if one chooses to represent quaternions via Pauli matrices, the real
		part of a quaternion exactly corresponds to the trace of its matrix representation.
	} ---
\begin{equation}
	a\, b\, c  ~~+~~  \text{q.c.}	~~=~~	b\, c\, a  ~~+~~  \text{q.c.}
\end{equation}
	The proof is simple --- the real part of a product of \emph{two} quaternions cannot depend
	on their order.
	From this follows the cyclicity.
	The fact that we are dealing with complexified quaternions cannot change this property.

	As we will see in Appendix~\ref{section-spinors}, in the product $ \psi_R\, \jj $ the factor $ \jj $
	``elevates'' the right-handed spinor $ \psi_R $ to the left-handed space
	(we have used the term ``elevates'' because conventionally a left-handed spinor is
	written above right-handed spinor inside the column of a Dirac spinor).
	Importantly, the factor $ \jj $ does not change the spinor's representation (it is obviously still transformed
	via multiplication by $ e^{\Lambda^*} $ on the left).
	Instead, the spinor just becomes expandable in the left-handed basis \eqref{lbasis}.
	We will use this when we define the \emph{standard} representation for spinors below.

	Written in terms of the components, the mass term \eqref{m-term} gives
\begin{equation}
	\psi_L^\dag\, \psi_R\, \jj  ~~+~~  \text{q.c.}  ~~+~~  \text{c.c.}	~~=~~
	\xi_L^*\,\xi_R  ~+~ \chi_L^*\,\chi_R  ~+~  \xi_R^*\,\xi_L  ~+~ \chi_R^*\,\chi_L\,,
\end{equation}
	as it should be for the Dirac mass term.

	We also mention here the mass term \eqref{m-term} can be written in an alternative form by
	replacing $ \jj $ with $ i\ii $,
\begin{equation}
\label{m-term-alt}
	i\ii\, m\, \psi_L^\dag\, \psi_R  ~~+~~  \text{q.c.}  ~~+~~  \text{c.c.}
\end{equation}
	Which form to use is a matter of convenience.

\subsection{Dirac Lagrangian}
	Now we can derive the Dirac equation.
	Our starting point is the full Lagrangian
\begin{equation}
\label{L-massive}
	\ml_\text{Dirac}	~~=~~	
			\psi_L^\dag\, i\md\, \psi_L  ~~+~~  \psi_R^\dag\, i\,\wt\md\, \psi_R
			~~-~~  
			\lgr m\, \psi_L^\dag\, \psi_R\, \jj  ~~+~~  \text{q.c.} \rgr  ~~+~~  \text{c.c.},
\end{equation}
	where we remember that the mass term has to actually enter with a negative sign,
	and we consider the electromagnetic field to be fixed.
	Having multiple terms, this Lagrangian can also be compacted into a form similar to \eqref{L-united}.
	We believe, however, that this should be automatically achieved in the context of the electroweak theory.
	For our current purposes, once again, it is important to establish the contact with Quantum Mechanics,
	and we prefer to keep the theory in the chiral form \eqref{L-massive}.

	To derive the equations of motion, we vary the Lagrangian \eqref{L-massive} with respect to
	$ \psi_L^\dag $ and $ \psi_R^\dag $.
	The reader may wonder at this point -- how are we going to differentiate this Lagrangian with respect
	to quaternions, let alone complexified and fermionic?
	We postpone the formal answer to this question until Appendix~\ref{section-diff}.
	For now, we can just act intuitively, at least when differentiating with respect to $ \psi_L^\dag $.
	Indeed, in the terms where it is present in Eq.~\eqref{L-massive}, it is sitting on the left,
	and so the na\"ive left derivative gives
\begin{equation}
\label{ldirac}
	i\,\md\,\psi_L  ~~-~~  m\, \psi_R\, \jj	~~=~~	0\,.
\end{equation}
	Speaking informally, for quaternions $ \partial \wt q/\partial q ~\neq~ 0 $ --- unlike for complex numbers,
	for which $ \partial \ov z / \partial z ~=~ 0 $ \cite{sudbery}, \cite{frenkel}.
	So it seems like there should be more terms on the left-hand side of Eq.~\eqref{ldirac}.
	Why we can act so na\"ively, and why the other terms do not contribute is, again,
	explained in Appendix~\ref{section-diff}.
	Here we provide an alternative and a more transparent justification, as follows.
	Let us for a moment pretend that we are dealing with a ``quaternionic'' Lagrangian
\begin{equation}
\label{L-quat}
	\psi_L^\dag\, i\md\, \psi_L  ~~+~~  m\, \psi_L^\dag\, \psi_R\, \jj
\end{equation}
	from which we will only need the real part.
	If we find the extremum of this Lagrangian, it will also extremize the real part of the latter.
	But the extremum of \eqref{L-quat} is exactly given by Eq.~\eqref{ldirac}.

	In order to vary with respect to $ \psi_R^\dag $, we just re-write the mass term as
\begin{equation}
	\ml_\text{Dirac}		~~\supset~~
			\lgr m\, \psi_R^\dag\, \psi_L\, \jj  ~~+~~  \text{q.c.} \rgr  ~~+~~  \text{c.c.}
\end{equation}
	We did not introduce anything new, as the first term here
	was just hidden inside the ``$ \text{q.c.} $'' and ``$ \text{c.c.} $'' in Eq.~\eqref{L-massive}.
	Notice that this term enters with a \emph{positive} sign.
	Now we can vary the Lagrangian with respect to $ \psi_R^\dag $, finding
\begin{equation}
\label{rdirac}
	i\,\wt\md\,\psi_R  ~~+~~  m\, \psi_L\, \jj	~~=~~	0\,.
\end{equation}
	The two expressions \eqref{ldirac} and \eqref{rdirac} are the \emph{quaternionic Dirac equations}.

\subsection{Non-relativistic limit}

	Now let us derive the Schr\"odinger equation.
	We do it in the classical way \cite{Berestetsky:1982aq}, by taking the large-mass limit.
	First, we open up the long derivatives in Eqs.~\eqref{ldirac}, \eqref{rdirac},
\begin{align}
\notag
	&
	i\,\p_0\,\psi_L  ~+~  A_0\,\psi_L  ~-~  (\vec\nabla ~+~ i\,\vec A)\,\psi_L
	~-~  m\, \psi_R\, \jj	~~=~~	0\,,
	\\[2mm]
\label{dirac-split}
	&
	i\,\p_0\,\psi_R  ~+~  A_0\,\psi_R  ~+~  (\vec\nabla ~+~ i\,\vec A)\,\psi_R
	~+~  m\, \psi_L\, \jj  ~~=~~ 0\,.
\end{align}

	We can make one consistency check.
	For a free particle, the time derivative gives the energy,
	$ i\,\p_0 ~\to~ E $.
	If at the same time, it has zero momentum, then $ E ~=~ m $, and
	we find
\begin{align}
\notag
	&
	m\,\psi_L  ~~-~~  m\,\psi_R\,\jj	~~=~~	0
	&
	\Longrightarrow\quad \psi_L	~~=~~	+\,\psi_R\,\jj\,,
	\\[2mm]
\label{rest}
	&
	m\,\psi_R  ~~+~~  m\,\psi_L\,\jj	~~=~~	0
	&
	\Longrightarrow\quad \psi_R	~~=~~	-\,\psi_L\,\jj\,.
\end{align}
	The two equations are consistent.
	What does the equality $ \psi_L ~=~ \psi_R\,\jj $ mean?
	As we already mentioned, and as we shall see in Appendix~\ref{section-spinors},
	multiplying by $ \jj $ on the right
	\emph{rotates} the components of $ \psi_L $ and $ \psi_R $ into
	each other.
	That is, in Dirac spinor language, for a spinor
\begin{equation}
	\psi_\text{chiral}	~~=~~	\lgr 
						\begin{matrix}
							\psi_L^\alpha	\\[2mm]
							\psi_{R\dot\alpha}
						\end{matrix}
					\rgr
\end{equation}
	this operation turns
\begin{align}
	&
	\psi_L^\alpha	~~\to~~		-\psi_{R\dot\alpha}\,,
	&
	\psi_{R\dot\alpha}	~~\to~~		\psi_L^\alpha\,.
\end{align}
	Equation \eqref{rest} then simply implies that the two spinors
	$ \psi_L^\alpha $ and $ \psi_{R\dot\alpha} $ are equal.
	This is correct, as a particle at rest is described by a single
	two-component spinor.

	This is too crude a limit, we want to keep the right-handed and left-handed
	spinors different to the order $ O(1/m) $, and momentum non-zero.
	
	To do that, we, essentially, switch to the standard representation.
	Let us multiply the second equation in \eqref{dirac-split} by $ \jj $
	on the right:
\begin{align}
\notag
	i\,\p_0\,\psi_L  ~+~  A_0\,\psi_L  ~-~  (\vec\nabla ~+~ i\,\vec A)\,\psi_L
	~-~  m\, \psi_R\, \jj	& ~~=~~	0\,,
	\\[2mm]
\label{dirac-j}
	i\,\p_0\,\psi_R\,\jj  ~+~  A_0\,\psi_R\,\jj  ~+~  (\vec\nabla ~+~ i\,\vec A)\,\psi_R\,\jj
	~-~  m\, \psi_L  & ~~=~~ 0\,.
\end{align}
	It is natural now to add and subtract these two equations, and introduce the notations
\begin{align}
	\psi_+		~~\equiv~~	\zeta	~~=~~	\frac{\psi_L \,+\, \psi_R \jj}{\sqrt{2}}\,,
	&&
	\psi_-		~~\equiv~~	\eta	~~=~~	\frac{\psi_L \,-\, \psi_R \jj}{\sqrt{2}}\,.
\label{standard}
\end{align}
	As discussed in Appendix~\ref{section-spinors}, these are precisely the upper
	and lower components of a Dirac spinor in the standard representation:
\begin{equation}
	\psi_\text{standard}	~~=~~	
		\lgr 
			\begin{matrix}
				\zeta	\\[2mm]
				\eta
			\end{matrix}
		\rgr.
\end{equation}
	In this section, we prefer to call them $ \psi_\pm $.
	For a particle at rest, $ \psi_- ~=~ 0 $ as it should be.
	Equations \eqref{dirac-j} now read
\begin{align}
\notag
	i\,\p_0\,\psi_+  ~+~  A_0\, \psi_+  ~-~  (\vec\nabla ~+~  i\,\vec A)\, \psi_-	& ~~=~~  \phantom{+}\,m\,\psi_+\,.
	\\[2mm]
\label{dirac-standard}
	i\,\p_0\,\psi_-  ~+~  A_0\, \psi_-  ~-~  (\vec\nabla ~+~  i\,\vec A)\,\psi_+	& ~~=~~ -\,m\,\psi_-\,.
\end{align}
	We now get rid of the mass term in the first equation in \eqref{dirac-standard} via the redefinition
\[
	\psi_\pm	~~\to~~		e^{-i m t}\, \psi_\pm\,.
\]
	The mass term is now gone from that equation:
\begin{equation}
	i\,\p_0\,\psi_+		~~=~~	(\vec\nabla ~+~ i\,\vec A)\,\psi_-  ~~-~~  A_0\, \psi_+\,,
\end{equation}
	while in the second equation the mass term doubles.
	Taking the large-mass limit we find
\[
	2m\,\psi_-	~~=~~	(\vec\nabla ~+~ i\,\vec A)\,\psi_+\,,
\]
	and so
\begin{equation}
	i\,\p_0\,\psi_+		~~=~~
		\frac 1 {2m}\, (\vec\nabla ~+~ i\,\vec A)^2\, \psi_+  ~~-~~  A_0\, \psi_+\,.
\end{equation}
	The final step now is to transform the quaternionic square
\[
	(\vec\nabla ~+~ i\,\vec A)^2
\]
	into the usual vector operators --- scalar and vector products.
	Denoting the scalar-product square of a vector by figure brackets $ \{~ \} $,
	one finds
\[
	(\vec\nabla ~+~ i\,\vec A)^2	~~=~~	-\, \big\{\vec\nabla ~+~ i\,\vec A\big\}^2  ~~+~~  i\,\vec B\,,
\]
	in the operator form.
	We thus get
\begin{equation}
\label{schrod}
	i\,\p_0\,\psi_+		~~=~~
		-\,\frac 1 {2m}\, \big\{ \vec\nabla ~+~ i\,\vec A\big \}^2\,\psi_+
		~~+~~  \frac{i\,\vec B}{2m}\, \psi_+  ~~-~~  A_0\,\psi_+\,.
\end{equation}
	Rescaling $ A^\mu ~\to~ e\,A^\mu $, and switching to the Gaussian units,
	we arrive to
\begin{equation}
\label{pauli}
	i\,\hbar\,\p_t\,\psi_+	~~=~~
		\lgr
			\frac 1 {2m}\, \big\{ \vec p ~+~ \frac e c\,\vec A \big\}^2
			~-~  e\,A_0
			~+~  \frac{i e \hbar}{2 m c}\,\vec B
		\rgr \psi_+\,,
\end{equation}
	which is nothing but the \emph{Pauli equation}.
	We again stress that the $ \sigma $-matrices are absent from equations \eqref{schrod} and \eqref{pauli}.
	Spin is the intrinsic property of the wavefunction.

	The factor of $ i $ in the magnetic moment term in Eq.~\eqref{pauli} has a double r\^ole.
	First, it is needed by the correspondence
\[
	\vec \sigma	~~\to~~		i\,\ii\,,~~ i\,\jj\,,~~ i\,\kk\,,
\]
	because the $ \sigma $-matrices are hermitean, and the quaternionic units are not.
	Second (and actually the same), by multiplying the magnetic field $ \vec B $,
	it turns it from an axial vector into a polar vector $ i\,\vec B $.
	In the regular spinor formalism, one yet has to argue that the product $ (\vec\sigma \cdot \vec B) $
	yields a true scalar when acting on $ \psi $.

	It is precisely the form of equation \eqref{pauli} that establishes both the quaternionic form
	of the Schr\"odinger equation
	and the fact that the Dirac Lagrangian \eqref{L-massive} is correct.
	Equation \eqref{pauli} involves a single two-component spinor $ \psi_+ $.
	As a complex quaternion, $ \psi_+ $ only occupies a half of the quaternionic space,
	and in the most	general form it can be written as
\[
	\psi_+		~~=~~	\psi\,P_L\,.
\]
	In fact this restriction can always be imposed in the end.
	The Schr\"odinger equation \eqref{pauli} can be solved for an arbitrary $ \psi $,
	and then projector $ P_L $ can be applied.
	The fact that the set of left-handed spinors forms an ideal guarantees that
	the projection will be a solution too.

	The right-handed projection $ \psi\,P_R $ will also be a solution, of course.
	It may either represent the same solution, or a different one,
	depending on the number of solutions with a given energy.
	It can always be ``moved'' to the left-handed space by multiplying by $ \jj $
	on the right.

	The magnetic moment term in Eq.~\eqref{pauli} is the only signature of quaternions
	in that equation.
	Without it, the equation looks exactly like the ordinary Schr\"odinger equation,
\begin{equation}
\label{simple-schrod}
	i\,\hbar\,\p_t\,\psi	~~=~~
		\lgr
			\frac {p^2} {2m}
			~+~  V
		\rgr \psi\,,
\end{equation}
	where we have dropped the gauge field for simplicity.
	For a complexified quaternion $ \psi $, this equation falls apart into four
	identical complex equations.
	So the non-relativistic Quantum Mechanics essentially only needs complex numbers.
	We stress again that it is the spin of the particle that calls for the
	quaternionic appearance of the wavefunction.

\subsection{Current}
\label{sec-current}

	Let us now find an expression for the electromagnetic current.
	It can be easily read off the Lagrangian.
	Indeed, the current by definition is whatever the gauge field couples to,
\begin{equation}
\label{L-current}
	\cell	~~\supset~~	j_\mu\, A^\mu	
		~~=~~	\frac{j\,\wt A ~+~ A\,\wt j } 2
		~~=~~	\frac{\wt A\, j  ~+~  \wt j\, A} 2\,.
\end{equation}
	We can differentiate (\emph{i.e.} vary) the latter quaternionic expression using equations
	\eqref{dq-vector} and \eqref{daq-vector}.
	The last fraction in Eq.~\eqref{L-current} suggests differentiating with respect to $ \wt A $ in order
	to get current $ j $,
\begin{equation}
	\p_{\wt A}\, \cell	~~\supset~~	\frac { 4\,j ~+~ (-2\,\wt{\wt j}) } 2
	~~=~~
	j\,.
\end{equation}
	We could now apply the same variational derivative to Lagrangian \eqref{L-massive}.
	Only this is not necessary --- it is enough to re-write the interacting part of
	\eqref{L-massive} in the form \eqref{L-current} using the cyclicity property,
\begin{align}
\notag
	\cell	& ~~\supset~~	\psi_L^\dag\, \wt A\, \psi_L  ~~+~~  \psi_R^\dag\, A\, \psi_R  ~~+~~
				\wt{\psi}{}_L\, A\, \psi_L^*  ~~-~~  \wt{\psi}{}_R\, \wt A\, \psi_R^*
	\\[3mm]
	& ~~=~~
	-\, \wt A\, \psi_L\,\psi_L^\dag  ~~+~~ \wt A\, \psi_R^*\,\wt{\psi}{}_R  ~~+~~  \text{q.c.},
\end{align}
	from which we conclude that
\begin{equation}
\label{current}
	j	~~=~~	-\,2 \lgr \psi_L\,\psi_L^\dag  ~-~  \psi_R^*\,\wt\psi{}_R \rgr.
\end{equation}
	We immediately observe that this expression transforms correctly under gauge transformations,
\begin{equation}
	j	~~\rightarrow~~	e^{i\varphi}\,j\,e^{-i\varphi}	~~=~~	j\,,
\end{equation}
	and under Lorentz transformations,
\begin{equation}
	j	~~\rightarrow~~	e^\Lambda\,j\,e^{\Lambda^\dag}\,.
\end{equation}
	Written explicitly in components, expression \eqref{current} gives,
\begin{align}
\notag
	j	&  ~~=~~			\big(\, \xi_L^*\,\xi_L ~+~ \chi_L^*\,\chi_L \,\big)  ~~+~~  
						\big(\, \xi_R^*\,\xi_R ~+~ \chi_R^*\,\chi_R \,\big)  ~~+~~
	\\[2mm]
\notag
		& ~~+~~		i\ii \lgr	\big(\, \xi_L^*\,\chi_L \,+\, \chi_L^*\,\xi_L \,\big) ~-~
						\big(\, \xi_R^*\,\chi_R \,+\, \chi_R^*\,\xi_R \,\big) \rgr
	\\[2mm]
		& ~~+~~		i\jj \lgr	i\,\big(\, \chi_L^*\,\xi_L \,-\, \xi_L^*\,\chi_L \,\big) ~-~
						i\,\big(\, \chi_R^*\,\xi_R \,-\, \xi_R^*\,\chi_R \,\big) \rgr
	\\[2mm]
		& ~~+~~		i\kk \lgr	\big(\, \xi_L^*\,\xi_L \,-\, \chi_L^*\,\chi_L \,\big) ~-~
						\big(\, \xi_R^*\,\xi_R \,-\, \chi_R^*\,\chi_R \,\big) \rgr,
\notag
\end{align}
	as it should be for the current written in the spinor representation.

	Finally, we can re-write current \eqref{current} using the standard representation \eqref{standard},
\begin{equation}
	j	~~=~~	-\, \lgr \zeta\,\zeta^\dag ~+~ \eta\,\eta^\dag ~+~ \text{c.c} \rgr  ~~-~~  
			\lgr \zeta\,\eta^\dag ~+~ \eta\,\zeta^\dag ~-~ \text{c.c} \rgr.
\end{equation}
	Here the first bracket gives the real part --- the $ j^0 $ component, while the second bracket gives
	the (imaginary) vector part.

\section{Maxwell's equation}
	Although not the main subject of our discussion, we can add the kinetic term for the gauge field,
	and derive the resulting Maxwell's equation.
	This section mainly serves the purpose of illustration of differentiating with respect to quaternionic vectors.
	One has to be particularly careful about the targets of differentiation of various involved derivatives.

	The gauge field strength is easily found to be
\begin{equation}
	\Phi	~~=~~	\vec B ~~+~~ i\,\vec E	~~=~~	-\, \frac{\p\,A \,-\, \wt A\,\wt\p} 2\,.
\end{equation}
	This expression becomes particularly simple in the Lorentz gauge:
\begin{equation}
	\Phi_\text{Lorentz}	~~=~~ -\, \p\, A\,,
\end{equation}
	although we will not use it for deriving the equations of motion.

	The relevant part of the Lagrangian looks as,
\begin{equation}
	\cell_\text{Maxwell}	~~=~~	\frac{ \frac 1 2\, \Phi^2 ~+~ \wt A\,j } 4  ~~+~~  \text{q.c.}  ~~+~~  \text{c.c.}
\end{equation}
	As we know from Section~\ref{sec-current}, varying the current part of the Lagrangian with respect to $ \wt A $
	expectably gives us the current,
\begin{equation}
	\p_{\wt A}\,\cell_\text{current}	~~=~~	\p_{\wt A}\, \frac{\wt A\,j \,+\, \wt j\, A} 2	~~=~~	j\,.
\end{equation}
	Now we need to vary the gauge part of the Lagrangian,
\begin{equation}
	\cell_\text{gauge}	~~=~~	\frac 1 4\, \Phi^2  ~~+~~  \text{c.c.}
\end{equation}
	
	Explicitly,
\begin{align}
\notag
	& \p_{\wt A}\, (\p\, A ~-~ \wt A\,\wt \p)\, (\p\, A ~-~ \wt A\,\wt \p)	~~=~~
	\\[3mm]
	& \uuline\p{}_{\wt A}\, (\p\, \uuline A ~-~ \wt{\uuline A}\,\wt \p)\, (\p\, A ~-~ \wt A\,\wt \p)
	~~+~~
	\uuline\p{}_{\wt A}\, (\p\, A ~-~ \wt A\,\wt \p)\, (\p\, \uuline A ~-~ \wt{\uuline A}\,\wt \p)\,.
\end{align}
	In the latter expression, we have underlined the particular instances of field $ A $ which are acted
	upon by $ \p_{\wt A} $ with a double line.
	All other occurrences of $ A $ are considered constant for the matter of variation.
	The space-time derivatives $ \p $ here act on the factors of $ A $ that are closest to them.
	We need to free the underlined variables $ A $ from the space-time derivatives, by integrating the latter
	by parts.
	This will change the sign in front of each of these derivatives, but will not move them anywhere
	because they are quaternionic.
	To indicate that they are now acting on different targets, we will underline them,
	as well as their targets, using a single line,
\begin{equation}
\label{by-parts}
	\uuline\p{}_{\wt A}\,	\big(\, -\,\uline\p\,\uuline A ~+~ \wt{\uuline A}\,\uline\p \,\big)
				\big(\, \uline{~~~~\dots~~~~} \,)
	~~+~~
	\uuline\p{}_{\wt A}\,	\big(\, \uline{~~~~\dots~~~~} \,)
				\big(\, -\,\uline\p\,\uuline A ~+~ \wt{\uuline A}\,\uline\p \,\big)\,.
\end{equation}
	Here the dots symbolize the term
\begin{equation}
\label{dots-term}
	\p\,A  ~~-~~  \wt A\,\wt\p\,.
\end{equation}
	For the sake of variation, the space-time derivatives are now just constants.
	In fact, everything is constant in Eq.~\eqref{by-parts} in regards to varying with respect to $ A $ ---
	except for the doubly-underlined $ A $ factors.
	Applying identities \eqref{dq-vector} and \eqref{daq-vector}, we get, explicitly
\begin{equation}
\label{varied}
	2\,\wt\p\, \big(\, ~~~\dots~~~ \,)  ~~+~~
	4\,\wt\p\, \big(\, ~~~\dots~~~ \,)  ~~+~~
	2\,\wt\p\, \big(\, ~~~\dots~~~ \,)\wt{~}  ~~+~~
	4\, \big(\, ~~~\dots~~~ \,)_0\,\wt\p\,.
\end{equation}
	The third term here involves the quaternionic conjugate of \eqref{dots-term},
	while the last term involves the real component of the \eqref{dots-term}.
	Since \eqref{dots-term} is purely imaginary, applying quaternionic conjugation just changes its sign,
	while taking the real part annihilates it,
\begin{equation}
	4\,\wt\p\,\big(\, \p\,A ~-~ \wt A\,\wt\p \,\big)	~~=~~	-\,8\,\wt\p\,\Phi\,.
\end{equation}

	Although a fair exercise in differentiation, the derivative of the conjugate term $ \ov\Phi{}^2 $
	will actually give the same contribution.
	The reason we can guess that is that the two terms $ \Phi^2 $ and $ \ov\Phi{}^2 $
	only differ by their complex-imaginary part
\begin{equation}
	\text{Im}~\Phi^2	~~\propto~~	\vec E \circ \vec B\,,
\end{equation}
	which is a boundary term and thus does not affect the equations of motion.

	Altogether, we arrive to,
\begin{equation}
	\p_{\wt A}\, \cell_\text{Maxwell}	~~=~~	j  ~~-~~  \wt\p\,\Phi\,,
\end{equation}
	or,
\begin{equation}
\label{Maxwell}
	\wt\p\,\Phi	~~=~~	j\,,
\end{equation}
	which is the \emph{quaternionic Maxwell's equation}.
	Notice that, because it is a complex-quaternionic equation, 
	it includes all eight real Maxwell's equations.
	The reader is encouraged to check that they are reproduced correctly in Eq.~\eqref{Maxwell}.

\section{Conclusions}

	We have demonstrated the construction of Dirac and Weyl spinors in the complex quaternionic
	space $ \cC \otimes \cH $.
	The spinors become part of the algebra, on the same grounds as the kinetic operators acting on them.
	This allows for finding purely-algebraic solutions of the Dirac equation in various settings,
	without resorting to the matrix form.
	Different types of conjugation of spinors \eqref{ferm-cc}, \eqref{ferm-qc} and \eqref{ferm-hc}
	are what particularly distinguishes the quaternionic formalism from the regular matrix description.
	For example, the discrete transformations \eqref{C-trans}--\eqref{T-trans} have a simpler
	appearance.
	In addition we notice that these complex and quaternionic conjugations are not as easy to implement
	in the regular spinor formalism.

	After building the Lagrangian, we have been able to develop a scheme of using the variational calculus
	to consistently derive the equations of motion.
	This calculus is well applicable to spinors and to vectors, which we have demonstrated by deriving the
	(single) Maxwell's equation.
	
	An important result is the digression of the algebra in the non-relativistic limit.
	By taking this limit, we have established the form of the Schr\"odinger's equation
	in the quaternionic formulation.
	Quantum Mechanics in the quaternionic form has been a subject of long study with various success.
	We show that it is possible to uniquely fix this form by starting from the Dirac equation.
	No extra degrees off freedom appear.
	Furthermore, the existence of the spin of the wavefunction is the natural consequence of the latter
	being quaternionic.
	The solution can be sought in the \emph{algebraic} form.
	Then, by applying the correct projector one finds the quantum-mechanical wavefunction.
	Such solutions will be the subject of further study.
	Once the spin is discarded, the Schr\"odinger's equation takes the usual \emph{complex}
	form.
	The quaternionic algebra reduces to the complex one in the non-relativistic limit.
	Extending this hypothesis, we hope to expect that addition of the strong and weak interactions
	grows the algebra to a unifying $ \cO \otimes \cH \otimes \cC \otimes \cR $.

	In this context, we can address the length of the Lagrangian of electrodynamics \eqref{L-massive}.
	As we have mentioned, it is possible to write it in a more compact way.
	There have been multiple reasons we have not done this in this paper.
	Besides the desired non-relativistic limit, we note that the Standard Model itself treats
	the chiralities differently.
	Finally, we believe, that the right answer to a compact form is given by placing the theory
	into the framework of electroweak interactions \cite{Furey:2015tqa}, \cite{thesis}.
	This comprises a promising direction for future work.

\section*{Acknowledgements}

	The author would like to thank Cohl Furey for valuable discussions during various stages
	of this work,
	and the Institute of Nuclear Theory at the University of Washington where part of this work
	was done for kind hospitality.

\pagebreak
\appendix
\setcounter{equation}{0}

\section{Spinors}
\label{section-spinors}

	How do we deduce the bases \eqref{lbasis} and \eqref{rbasis} for spinors?
	In this section we essentially expand on the construction introduced in \cite{thesis}.
	We start from the correspondence
\begin{equation}
	\sigma_1\,,~~ \sigma_2\,,~~ \sigma_3	~~\to~~		i\,\ii\,,~~ i\,\jj\,,~~ i\,\kk\,,
\end{equation}
	for the Pauli matrices.
	This is a very natural association, and the only other reasonably alternative choice here
	would be a different sign on the right-hand side.

        Next we build the spin-up and spin-down states, assuming the rest frame of reference.
	The operator of the canonical $ z $-component of spin should be
\[
	\frac{\sigma^3} 2	~~\to~~		\frac{i\,\kk} 2\,,	
\]
	so the up- and down-states should satisfy
\begin{equation}
	i\,\kk\, \psi_\uparrow	~~=~~		+\,\psi_\uparrow
\end{equation}
	and
\begin{equation}
	i\,\kk\, \psi_\downarrow	~~=~~		-\,\psi_\downarrow\,.
\end{equation}
	Or, in other words,
\begin{equation}
\label{updown}
	(1 ~\mp~ i\,\kk)\, \psi_{\uparrow,\downarrow}	~~=~~	0\,.
\end{equation}

	The reason that a product of two complex quaternions can vanish, is because
	the algebra of complex quaternions does not admit a positive-definite norm.
	We can still use the usual quaternionic norm,
\begin{equation}
\label{qnorm}
	\|\,a\,\|^2	~~=~~	a_0^2  ~~+~~  a_1^2  ~~+~~  a_2^2  ~~+~~  a_3^2\,,
\end{equation}
	ignoring the fact that $ a_0 $, ... are complex numbers.
	Such a norm will be multiplicative, but not positive-definite.
	The reason the product of the two factors in Eq.~\eqref{updown} vanishes 
	is because the norm of at least one of those factors vanishes.
	Indeed, the norm of $ 1 ~\mp~ i\,\kk $ as calculated via Eq.~\eqref{qnorm} is zero.

	So how about $ \psi_{\uparrow,\downarrow} $? How many solutions can there be?
	These questions are easily answered by noticing that objects $ 1 ~+~ i\,\kk $ 
	and $ 1 ~-~ i\,\kk $ are, in fact, projectors:
\begin{align}
	\notag
	&
	P_L	~~=~~	\frac { 1 ~+~ i\,\kk } 2\,,
	&
	P_L^2	~~=~~	P_L\,,
	\\[3mm]
	&
	P_R	~~=~~	\frac { 1 ~-~ i\,\kk } 2\,,
	&
	P_R^2	~~=~~	P_R\,,
\end{align}
	which we judiciously have named the left- and right-handed projectors.
	Indeed, by multiplying a quaternion on the left (or equally well, on the right) by $ P_L $,
	we are obviously performing a linear operation upon the components of that quaternion.
	The square of such an operation equals the operation itself $ \,\Rightarrow\, $
	the operation is a projection.
	Obviously the same is true for $ P_R $.

	That means that if we run $ \psi $ through all values of complex quaternions,
	the product $ P_L \psi $ will span only a portion of the quaternionic space
	--- an ideal.
	What fraction of the entire algebra does it span?
	We notice that $ P_L $ and $ P_R $ are complementary projectors:
\begin{align}
	&
	P_R	~~=~~	P_L^*\,,
	&
	P_L  ~~+~~  P_R		~~=~~	1\,.
\end{align}
	Because they are symmetric, they can only project equal-size subsets of the algebra.
	And since they are complementary to each other, the union of those subsets
	must comprise the entire algebra.
	In other words, $ P_L $ and $ P_R $ split the algebra in two halves.

	Now, equation \eqref{updown} can re-written as,
\begin{equation}
\label{PRL}
	P_{R,L}\, \psi_{\uparrow,\downarrow}		~~=~~		0\,,
\end{equation}
	meaning that the most general $ \psi_\uparrow $ must sit in one half of the algebra,
	while the most general $ \psi_\downarrow $ must sit in the other half.
	Since a generic complex quaternion has four complex components,
	there are two complex solutions for $ \psi_\uparrow $ and as many for $ \psi_\downarrow $.
	We have to stress here that these halves are \emph{not identified} with the left-
	and right-handed chiral spaces.
	For this reason we have not given names to these subspaces, other than ``spin-up'' and
	``spin-down'' spaces.

	The easiest solution to \eqref{PRL} is given by the orthogonality of the projectors:
\begin{align}
	&
	P_R\, P_L	~~=~~	0\,,
	&
	P_L\, P_R	~~=~~	0\,.
\end{align}
	So, seemingly $ P_L $ could be identified with a spin-up state, and $ P_R $ --- with the corresponding
	spin-down state.
	However, these are states of different chiralities.
	Indeed, since $ P_L ~=~ P_R^* $, such ``spinors'' transform in the mutually-conjugate
	representations of the Lorentz group.
	This fact puts them into the opposite chirality spaces.

	Let us summarize our goal and achievements now.
	We are looking for four complex states $ \psi_{L\uparrow} $, $ \psi_{L\downarrow} $, $ \psi_{R\uparrow} $
	and $ \psi_{R\downarrow} $.
	We have already found
\begin{align}
	&
	\psi_{L\uparrow}		~~=~~	P_L\,,
	&
	\psi_{R\downarrow}	~~=~~	P_R\,.
\end{align}
	It is not difficult to find the other two states.
	For example, if one wishes to avoid pure guessing, which would perfectly work here too,
	we know that matrix $ -i \sigma^2 $ turns a spin-up state into a spin-down state:
\[
	-i\,\sigma^2\,
		\lgr
		\begin{matrix}
			1 \\[2mm]
			0
		\end{matrix}
		\rgr
	~~=~~
		\lgr
		\begin{matrix}
			0	&	-1 \\[2mm]
			1	&	0
		\end{matrix}
		\rgr
		\lgr
		\begin{matrix}
			1 \\[2mm]
			0
		\end{matrix}
		\rgr
	~~=~~
		\lgr
		\begin{matrix}
			0 \\[2mm]
			1
		\end{matrix}
		\rgr.	
\]
	This exactly corresponds to multiplying by $ \jj $ on the left, and we arrive to
\begin{align}
	&
	\psi_{L\downarrow}	~~=~~	\phantom{-}\jj\,P_L\,,
	&
	\psi_{R\uparrow}		~~=~~	-\jj\,P_R\,.
\end{align}

	So we recap that the \emph{left-handed} spinors can be written as
\begin{equation}
\label{lbasis-again}
	\psi_L	~~=~~	\xi_L\,P_L  ~~+~~  \chi_L\,\jj\,P_L\,,
\end{equation}
	while the \emph{right-handed} spinors are represented as
\begin{equation}
\label{rbasis-again}
	\psi_R	~~=~~	-\xi_R\,\jj\,P_R  ~~+~~  \chi_R\,P_R\,.
\end{equation}

	A few important comments are in order here.
	The chiral subspaces are defined here by multiplying
	arbitrary quaternions by projectors $ P_L $ or $ P_R $ on the \emph{right}.
	In other words, $ \psi\, P_L $ spans the set of all left-handed spinors,
	and $ \psi\, P_R $ of all right-handed.
	This way, right multiplication splits the set of quaternions into
	the two chirality subspaces.
	Whereas, \emph{left} multiplication splits the quaternions into
	spin-up and spin-down subspaces, because \emph{e.g.}
\[
	P_R\, (P_L\, \psi)	~~=~~	0\,.
\]
	It is a trivial fact now that a set of the type $ \big\{\, P_L\, \psi \,\} $ or $ \big\{\, \psi\, P_L \,\} $
	forms a left (right) ideal, since for an arbitrary quaternion $ a $,
	the product, say,
\[
	a\, (\psi\, P_L)	~~=~~	(a\, \psi)\, P_L\,,
\]
	resides in the same subspace as $ \psi\, P_L $.

	The other important remark is about the amount of freedom that we have in defining
	the bases via Eqs.~\eqref{lbasis-again}, \eqref{rbasis-again}.
	We have freedom in choosing the component of the spin to be measurable --- for which we chose $ \kk $.
	The other freedom was in parametrizing the spin-down component of $ \psi_L $,
	for which we chose $ \jj $ as an orthogonal direction.
	Overall, we could have chosen any two unit vectors $ \hat a $ and $ \hat b $ in place of $ \kk $
	and $ \jj $, subject to the only restriction
\[
	\hat a \cdot \hat b	~~=~~	0\,.
\]

	The \emph{standard representation} for the spinors can be established as follows.
	We notice that multiplying a left-handed spinor $ \psi_L $ on the right by $ \jj $
	brings it into the right-handed subspace,
\begin{equation}
	\psi_L\,\jj	~~\sim~~	-\,\psi_R\,,
\end{equation}
	and, respectively, a right-handed spinor $ \psi_R $ into the left-handed space,
\begin{equation}
	\psi_R\,\jj	~~\sim~~	\psi_L\,.
\end{equation}
	This way, by forming a linear combination
\begin{equation}
	\frac{\psi_D  ~+~  \psi_D\,\jj}{\sqrt 2}	~~=~~	\psi_D\, \frac{ 1 ~+~ \jj }{\sqrt 2}\,,
\end{equation}
	we create a spinor consisting of
\begin{equation}
	\zeta	~~=~~	\frac{\psi_L \,+\, \psi_R}{\sqrt 2}
	\qquad\qquad\text{and}\qquad\qquad
	-\,\eta	~~=~~	\frac{\psi_R \,-\, \psi_L}{\sqrt 2}\,.
\end{equation}
	Here the sum and difference are understood in terms of the usual spinor components.
	In terms of the quaternionic addition, spinors $ \zeta $ and $ \eta $
	can be written \emph{e.g.} as
\begin{align}
	\zeta	~~=~~	\frac{\psi_L \,+\, \psi_R \jj}{\sqrt{2}}\,,
	&&
	\eta	~~=~~	\frac{\psi_L \,-\, \psi_R \jj}{\sqrt{2}}\,.
\end{align}
	In this example they both reside in the left-handed subspace --- for definiteness,
	but this has no special r\^ole in the standard representation,
	and the corresponding right-handed expressions can be readily written.

	As was discussed in \cite{thesis}, the Dirac algebra
	C$\ell$(4) $ \sim $  C$\ell$(2) $ \otimes $ C$\ell$(2)
	is realized in a very interesting way on complex quaternions.
	One factor of C$\ell$(2) acts on $ \psi_D $ via multiplication on the \emph{left}
	while the other one --- via multiplication on the \emph{right}.
	Left multiplication by quaternions rotates the spin components in 
	$ \psi_L $ and $ \psi_R $ independently, while right multiplication
	does not rotate the spin components, and instead rotates 
	$ \psi_L \,\leftrightarrow\, \psi_R $.
	This latter rotation can be used to identify the discrete symmetries.

\subsection{Discrete symmetries}
\label{section-discrete}

	Here we will discuss the action of $ C $, $ P $ and $ T $ symmetries on fermions.
	These symmetries can be conveniently written for an entire Dirac fermion $ \psi_D $.
	We use here the phase conventions of \cite{Berestetsky:1982aq}.

\begin{itemize}

\item
	Charge conjugation $ C $ is realized by complex conjugation of the fermion,
\begin{equation}
\label{C-trans}
	\psi_D(x)	~~\to~~		i\,\psi_D^*(x)\,.
\end{equation}

\item
	Parity transformation $ P $ is given by right multiplication by $ -\ii $,
\begin{equation}
\label{P-trans}
	\psi_D(t,\, \vec x)	~~\to~~		-\,\psi_D(t,\, -\vec x)\,\ii\,,
\end{equation}
	which interchanges the chiral components as,
\begin{equation}
	\psi_L	~~\to~~		i\,\psi_R\,,
	\qquad\qquad
	\psi_R	~~\to~~		i\,\psi_L\,.
\end{equation}

\item
	Time inversion $ T $ is performed by complex conjugation and 
	multiplying by $ i\jj $ on the right,
\begin{equation}
\label{T-trans}
	\psi_D(t,\, \vec x)	~~\to~~		i\,\psi_D^*(-t,\, \vec x)\,\jj\,.
\end{equation}

\end{itemize}

	All three transformations result in
\begin{equation}
\label{CPT-trans}
	C\,P\,T\, \psi_D(x)	~~=~~		\psi_D(-x)\,\kk\,.
\end{equation}
	Right-multiplying by $ i\kk $ preserves the sign of the left-handed spinor,
	while flips the sign of the right-handed part.
	Therefore, multiplying by $ \kk ~=~ (-i)i\kk $ corresponds to multiplying
	by $ i\gamma^5 $ in the spinor representation.
	Notice that the above transformations have a significantly simpler
	form than they do in the $ \gamma $-matrix representation.

\section{Differentiation}
\label{section-diff}

	Derivation of the equations of motion from the Lagrangian involves one
	crucial step --- variation, which we will loosely call differentiation.
	Indeed, the problem of extremizing the action essentially reduces
	to differentiating with respect to a quaternion.
	As soon as we are able to differentiate, variational calculus sets in place.
	We start with regular quaternions.

\subsection{Ordinary quaternions}
\label{sec-ordinary-quat}

	It is well known that there is no existing analyticity theory of quaternions
	analogous to that of complex numbers.
	Even the notion of a derivative is not well established.
	While complex analyticity rests on holomorphic functions that depend on $ z $ and
	are independent of $ \ov z $,
\[
	f(z,\, \ov z)	~~=~~	f(z)\,,
\]
	this property cannot be extended to quaternions.

	There is one crucial reason for this: if a complex function depends on $ z $, there is no
	way that it can be represented as a function of $ \ov z $.
	In other words, variable $ z $ cannot be converted into $ \ov z $ by
	multiplying it by any constants.
	This is \emph{not so} with quaternions.
	For, given a quaternion $ q $, we find that a combination
\begin{equation}
	-\, \frac{ q  ~+~  \ii\,q\,\ii  ~+~  \jj\,q\,\jj  ~+~  \kk\,q\,\kk } 2
	~~=~~
	\wt q
\end{equation}
	is exactly the conjugate of $ q $.
	Any other type of conjugation that can be introduced for quaternions (say, the one that
	only flips the sign of a single imaginary unit $ \ii $) can also be represented in such
	an ``arithmetic'' form.
	That means that any function $ f(q) $ can be viewed as a function of $ \wt q $, just
	with different coefficients.
	So the notion of holomorphy cannot be applied to quaternions, at least directly.

	It is, however, possible to define a convenient notion of a derivative.
	It works especially well if the function being differentiated is real (although it does not have to be).

	Let us say that $ f(q) $ is a function of variable $ q $,
\begin{equation}
	q	~~\equiv~~	t  ~~+~~  \vec r	~~=~~	t  ~~+~~  x\,\ii  ~~+~~  y\,\jj  ~~+~~  z\,\kk.
\end{equation}
	We introduce a derivative $ \p $,
\begin{equation}
\label{diff-d}
	\p	~~\equiv~~   \p_t  ~~+~~  \vec\nabla	~~=~~	\p_t  ~~+~~  
								\ii\,\p_x  ~~+~~  \jj\,\p_y  ~~+~~  \kk\,\p_z\,.
\end{equation}
	Strictly speaking, in analogy to complex numbers,
	we should call this quantity a conjugate derivative $ 2\,\wt\p $ multiplied by a factor of two or so,
	as we will see below. However, to keep notations flat and straightforward, we just
	call it $ \p $, and we define a conjugate derivative as
\begin{equation}
\label{diff-d-conj}
	\wt\p	~~\equiv~~   \p_t  ~~-~~  \vec\nabla	~~=~~	\p_t  ~~-~~  
								\ii\,\p_x  ~~-~~  \jj\,\p_y  ~~-~~  \kk\,\p_z\,.
\end{equation}

	Let $ a $ denote an arbitrary quaternionic constant.
	Then the following key identities
\begin{align}
\notag
	\p\,q		& ~~=~~		-2		& \p\,\wt q	& ~~=~~		~~~~4\\[2mm]
\label{dq}
	\wt\p\,q	& ~~=~~		~~~~4		& \wt\p\,\wt q	& ~~=~~		-2\,,
\end{align}
	and
\begin{align}
\notag
	\p\, (a q)	& ~~=~~		-2\,\wt a	& \p\, (a \wt q)	& ~~=~~		~~~~4\,a_0\\[3mm]
\label{daq}
	\wt\p\, (a q)	& ~~=~~		~~~~4\,a_0	& \wt\p\, (a \wt q)	& ~~=~~		-2\,\wt a\,.
\end{align}
	enable us to perform differentiation.
	Note that the set of identities \eqref{dq} applies equally well when the derivative is acting from the right.

	Any function of $ q $ which can be represented as a power series (in fact, any \emph{analytical} function
	of coordinates $ t $, $ x $, $ y $, $ z $), 
\begin{equation}
\label{q-series}
	f(q)	~~=~~	\dots ~~+~~ \alpha\,q\,\beta\,q\,\gamma\,q\,\delta  ~~+~~ \dots
\end{equation}
	can now be differentiated using Eqs.~\eqref{dq}, \eqref{daq}.
	It is sufficient to define the action of the derivative\footnote{We are not really ``defining'' this action,
	as it is explicitly defined in Eq.~\eqref{diff-d}, but instead giving a recipe how to efficiently compute such a derivative.}
	on the monomial in \eqref{q-series}.
	Derivative $ \p $ will act on each factor of $ q $ in \eqref{q-series} in turn.
	Let us underline each factor when it is differentiated, and when it is not --- it can be considered a constant:
\begin{equation}
\label{diff-term}
	\p~ \alpha\,q\,\beta\,q\,\gamma\,q\,\delta	~~=~~
		\underline \p~ \alpha\,\underline q\,\beta\,q\,\gamma\,q\,\delta  ~~+~~
		\underline \p~ \alpha\,q\,\beta\,\underline q\,\gamma\,q\,\delta  ~~+~~
		\underline \p~ \alpha\,q\,\beta\,q\,\gamma\,\underline q\,\delta\,.
\end{equation}
	We are also underlining the derivative operator, so that in the case when there is more than
	one derivative is present, it is clear which one acts on what variable.
	Every factor in \eqref{diff-term} that is not differentiated can be considered a constant.
	Thus, we can use the first one of identities \eqref{daq},
\[
	\p~ \alpha\,q\,\beta\,q\,\gamma\,q\,\delta	~~=~~
		-\,2\,\wt\alpha\,\beta\,q\,\gamma\,q\,\delta  ~~-~~
		2\, \wt\beta\, \wt q\, \wt\alpha\, \gamma\, q\, \delta  ~~-~~
		2\, \wt\gamma\, \wt q\, \wt\beta\, \wt q\, \wt\alpha\, \delta\,.
\]
	The efficiency of this approach is that we did not have to deal with components.

	In which sense is $ \p f(q) $ a derivative of $ f(q) $?
	Can we restore an infinitesimal change $ \Delta f $ due the increment $ \Delta q $, using this derivative?
	The answer is not quite straightforward, as the correct expression is given by
\begin{equation}
\label{delta-f}
	\Delta f	~~=~~	\underline f(q)\, \frac{ \wt\p\, \Delta q ~+~ \Delta \wt q\, \p } 2\,.
\end{equation}
	This result is obvious because the fraction here actually equals $ \p_\mu\, \Delta q^\mu $.
	Being a scalar, it can be written on either side of $ f(q) $.
	The problem here is that one of the two terms in that fraction will
	necessarily have a derivative operator $ \p $ sitting furthest away from $ f(q) $,
\[
	f(q)\, \Delta \wt q\, \p\,,
\]
	and so will not form a derivative in our sense.
	In other words, here the derivative has to depend on the increment $ \Delta q $
	in order to correctly reproduce $ \Delta f $.

	Instead of engaging with this problem, let us switch to \emph{real} functions $ f(q) $ ---
	our prime subject of interest --- since the Lagrangian is real.
	For real $ f(q) $, the rightmost term in the fraction in \eqref{delta-f} commutes with $ f(q) $,
	and so
\[
	\Delta f	~~=~~	\frac 1 2 \lgr \Delta \wt q\, \p\, f(q)  ~~+~~  f(q)\, \wt \p\, \Delta q \rgr,
\]
	where we also have replaced $ f(q) $ with $ \wt f(q) $ in the first term in the bracket.
	The extremum of $ f(q) $ is obviously achieved when
\begin{equation}
	\p\,f(q)	~~=~~	0\,,
\end{equation}
	since this also implies that $ \wt f(q)\, \wt \p ~=~ 0 $.

	Let us see how this works in practice.
	Let us apply a derivative operator $ \p $ on a real function $ f(q) $.
	Even if we do not know the form of $ f(q) $, when viewed as a series it can always be represented as
\begin{equation}
\label{f-real}
	f(q)		~~=~~	\dots  ~~+~~  q\,a  ~~+~~  \wt a\, \wt q  ~~+~~  \dots\,,
\end{equation}
	since when we differentiate a particular factor of $ q $ in a given monomial term,
	all the other factors of $ q $ in that term can be considered constant and absorbed into $ a $.
	Applying identities \eqref{dq}, \eqref{daq} to function \eqref{f-real}, we get
\begin{equation}
\label{der-f}
	\p\, f(q)	~~\supset~~	-2\,a  ~~+~~  4\,a_0	~~=~~	2\,\wt a\,.
\end{equation}
	Note that we would have arrived to the same result had we written $ q $ and $ a $ in Eq.~\eqref{f-real}
	in a different order,
\begin{equation}
	f(q)		~~=~~	\dots  ~~+~~  a\,q  ~~+~~  \wt q\, \wt a  ~~+~~  \dots\,,
\end{equation}
	only now the other identities in \eqref{dq}, \eqref{daq} would have been involved.

	Now let us apply operator $ \wt\p $ to Eq.~\eqref{f-real}.
	We find,
\begin{equation}
\label{der-f-conj}
	\wt \p\, f(q)	~~=~~	\dots  ~~+~~  4\,a  ~~-~~  2\,a  ~~+~~  \dots	
			~~=~~	\dots  ~~+~~  2\,a  ~~+~~  \dots\,.
\end{equation}
	Equations \eqref{der-f} and \eqref{der-f-conj} tell us that $ (1/2)\,\p $ in fact differentiates
	with respect $ \wt q $, while $ (1/2)\,\wt \p $ --- with respect to $ q $.
	This is exactly the same as with complex numbers, even with a matching factor of $ 1/2 $.
	Still, we would like to leave our definitions of $ \p $ and $ \wt \p $ as they are
	in order to keep notations plain.

	Being able to find the extremum of a real function, we now know how to derive the
	extremum of the action --- that is, the equations of motion.
	The derivative gets promoted to a variational derivative.

\subsection{Generic complex quaternions}
\label{sec-cpx-quat}
	
	Does the above picture of differentiation change for complex quaternions?
	If we are talking about generic complex quaternions --- not really.
	Generic complex quaternions have eight real degrees of freedom.
	Everything stays the same, taking into account that the components $ t $, $ x $, $ y $ 
	and $ z $ are now \emph{complex numbers}.

	The derivatives
\begin{align}
\notag
	\p	& ~~=~~	\p_t  ~~+~~  \ii\,\p_x  ~~+~~  \jj\,\p_y  ~~+~~  \kk\,\p_z\,,
	\\[2mm]
\label{der-cquat}
	\wt\p	& ~~=~~	\p_t  ~~-~~  \ii\,\p_x  ~~-~~  \jj\,\p_y  ~~-~~  \kk\,\p_z
\end{align}
	stay exactly what they are.
	Here, of course, we assume that the complex derivatives are defined in the usual way,
\begin{equation}
	\p_t	~~=~~	\frac 1 2\, \big\{\, \p_{t_1}  ~-~ i\,\p_{t_2}\,\big\} \,,\qquad	\dots\,.
\end{equation}
	We also have the two conjugates of \eqref{der-cquat},
\begin{align}
\notag
	\p^*		& ~~=~~	\p_t^*  ~~+~~  \ii\,\p_x^*  ~~+~~  \jj\,\p_y^*  ~~+~~  \kk\,\p_z^*\,,
	\\[2mm]
\label{der-cquat-conj}
	\p^\dag	& ~~=~~	\p_t^*  ~~-~~  \ii\,\p_x^*  ~~-~~  \jj\,\p_y^*  ~~-~~  \kk\,\p_z^*.
\end{align}
	Obviously, any function of $ q $ (and not $ q^* $) will have 
	derivatives $ \p^* $ and $ \p^\dag $ vanishing on it.

	Relations \eqref{dq} and \eqref{daq} are valid in our complex case
	verbatim.
	There is also a complex conjugated copy of them.
	We are not going to pay much attention to these relations.

	We are able to re-use literally all the formulas applicable to the differentiation
	with respect to regular quaternions, because complex quaternions have all
	eight degrees of freedom occupied.

	The story becomes more interesting when we have to address \emph{constrained}
	quaternions --- \emph{e.g.} vectors or spinors.
	We call them constrained, because not all eight components in them are independent
	(or non-zero).
	Thus, the derivatives \eqref{der-cquat}, \eqref{der-cquat-conj} cannot be applied to
	them verbatim to produce a meaningful result.

	There is also the field strength $ \Phi $ which is a constrained quaternion,
	but we normally do not vary with respect field strengths.

\subsection{Constrained complex quaternions --- vectors}

	Vectors open up quite an interesting story, as the only physical vector
	field variables are gauge fields.
	Thus, differentiation with respect to vectors enables us to derive
	an expression for the current, as well as to formally derive the form 
	of the quaternionic Maxwell's equation.

	Formally vectors can be defined by the constraint condition
\begin{equation}
	q^\dag	~~=~~	q\,,
\end{equation}
	which means that only four real components of it are non-zero:
\begin{equation}
	q	~~\equiv~~	t  ~~+~~  i\,\vec r
		~~=~~		t  ~~+~~  i\,\big(\, \ii\,x  ~~+~~  \jj\,y  ~~+~~  \kk\,z \,\big)\,.
\end{equation}

	In relation to it we define the derivative operators
\begin{equation}
	\p	~~\equiv~~	\p_t  ~~+~~  i\,\vec\nabla
		~~=~~		\p_t  ~~+~~  i\,\big( \ii\,\p_x  ~~+~~  \ii\,\p_y  ~~+~~  \kk\,\p_z \big)
\end{equation}
	and
\begin{equation}
	\wt\p	~~=~~	\p^*
		~~=~~	\p_t  ~~-~~  i\,\big( \ii\,\p_x  ~~+~~  \ii\,\p_y  ~~+~~  \kk\,\p_z \big)\,.
\end{equation}

	Relations \eqref{dq} and \eqref{daq} are valid but now their right-hand sides are interchanged vertically,
\begin{align}
\notag
	\p\,q		& ~~=~~		~~~~4		& \p\,\wt q	& ~~=~~		-2\\[2mm]
\label{dq-vector}
	\wt\p\,q	& ~~=~~		-2		& \wt\p\,\wt q	& ~~=~~		~~~~4\,,
\end{align}
	and
\begin{align}
\notag
	\p\,(a q)	& ~~=~~		~~~~4\,a_0	& \p\,(a \wt q)		& ~~=~~		-2\,\wt a\\[3mm]
\label{daq-vector}
	\wt\p\, (a q)	& ~~=~~		-2\,\wt a	& \wt\p\, (a \wt q)	& ~~=~~		~~~~4\,a_0\,.
\end{align}
	Other than that, there is no difference from differentiating with respect to regular quaternions,
	and these formulas are enough to derive the equations of motion.

\subsection{Constrained complex quaternions --- spinors}

	Full Dirac spinors are just complex quaternions, and so differentiation
	with respect to them is in fact addressed by operators \eqref{der-cquat}.
	There is nothing new from that perspective.
	One only has to keep in mind that spinors are Grassmann numbers.

	Weyl spinors, however are constrained.
	Even though a single Weyl spinor $ \psi_L $ or $ \psi_R $ has all eight
	components non-zero, it only has four independent ones.

	Let us take a left-handed fermion $ \psi_L $,
\begin{equation}
	\psi_L	~~=~~	\xi_L\,P_L  ~~+~~  \chi_L\,\jj\,P_L\,.
\end{equation}
	It turns out, that differentiation with respect to $ \psi_L $ is only
	meaningful when done from the right.
	The reason behind is that when multiplying by anything from the left, it is 
	not possible to ``cancel'' the projectors $ P_L $.
	We have,
\begin{equation}
\label{diff-psiL}
	\underline{\psi_L}\, \lgr \p_{\xi_L} ~-~ \jj\, \p_{\chi_L} \rgr
	~~=~~
	P_L  ~~+~~  P_R
	~~=~~
	1\,.
\end{equation}
	Again, we have underlined $ \psi_L $ to show that the derivatives act to the left.
	Here $ \p_{\xi_L} $ and $ \p_{\chi_L} $ are the usual Grassmann derivatives with respect to
	the complex components $ \xi_L $ and $ \chi_L $ correspondingly.
	Analogously we find
\begin{equation}
\label{diff-psiR}
	\underline{\psi_R}\, \lgr \jj\, \p_{\xi_R} ~+~ \p_{\chi_R} \rgr	~~=~~
	\underline{\psi_R}\, \jj\! \lgr \p_{\xi_R} ~-~ \jj\, \p_{\chi_R} \rgr	~~=~~
	1\,.
\end{equation}
	So we essentially have identified the spinor derivatives, we just need to re-write them
	in terms of the quaternionic components $ t $, $ x $, $ y $ and $ z $.
	Although we are really considering $ \psi_L $ and $ \psi_R $ separately,
	let us combine them both into a single Dirac fermion.
	This is only done for the sake of completeness --- a derivative with respect to $ \psi_L $
	does not touch $ \psi_R $ in any sense, 
	because the components of the two spinors are independent complex numbers.
	We have,
\begin{align}
\notag
	\psi_D	& ~~=~~
	\frac{ \xi_L \,+\, \chi_R } 2  ~~+~~
	\ii\, i\,\frac{\xi_R \,+\, \chi_L} 2  ~~+~~
	\jj\, \frac{-\xi_R \,+\, \chi_L} 2  ~~+~~
	\kk\, i\,\frac{\xi_L \,-\, \chi_R} 2
	\\[4mm]
\label{psiD-as-quat}
	& ~~=~~
	t  ~~+~~  \ii\, x  ~~+~~  \jj\, y  ~~+~~  \kk\, z\,.
\end{align}
	Knowing this, we can express derivatives $ \p_{\xi_L}, $ \emph{etc} in terms
	of the derivatives with respect to components $ t $, $ x $, $ y $ and $ z $.
	Plugging them into the expressions in parentheses in equations
	\eqref{diff-psiL} and \eqref{diff-psiR} we find,
\begin{align}
\notag
	\p_{\psi_L}	& ~~\propto~~
	\frac{\p_t ~-~ i\jj\,\p_x ~-~ \jj\,\p_y ~+~ i\,\p_z} 2\,,
	\\[5mm]
\label{prelim-psi-der}
	\p_{\psi_R}	& ~~\propto~~
	\frac{\p_t ~+~ i\jj\,\p_x ~-~ \jj\,\p_y ~-~ i\,\p_z} 2\,.
\end{align}
	Although we could write equality signs in these equations, we postpone this
	until we do the last extra step.
	Namely, equations \eqref{prelim-psi-der} do not seem to be very aesthetic,
	or easy to remember.
	Notice, that the action of these derivatives will not change if we multiply
	them by the projectors $ P_L $ and $ P_R $ on the left, correspondingly.
	Indeed, these projectors will carve out $ \psi_L $ or $ \psi_R $ from Eq.~\eqref{psiD-as-quat},
	before the actual derivatives will hit them.
	It is then a matter of a simple multiplication to observe that the right-hand
	sides of Eq.~\eqref{prelim-psi-der} when multiplied by the corresponding projectors,
	match a regular derivative $ \wt\p $ multiplied by the same projectors
\begin{align}
\notag
	\p_{\psi_L}	& ~~=~~		\frac 1 2\, P_L\, \wt\p\,,
	\\[4mm]
\label{spinor-der}
	\p_{\psi_R}	& ~~=~~		\frac 1 2\, P_R\, \wt\p\,.
\end{align}
	where $ \wt\p $ is the derivative \eqref{der-cquat} with respect
	to the components \eqref{psiD-as-quat} of $ \psi_D $ viewed as a complex quaternion,
\begin{equation}
	\wt\p	~~=~~	\p_t  ~~-~~  \ii\,\p_x  ~~-~~  \jj\,\p_y  ~~-~~  \kk\,\p_z\,.
\end{equation}
	From Eq.~\eqref{spinor-der}, 
\begin{equation}
\label{spinor-der-total}
	\p_{\psi_L}  ~~+~~  \p_{\psi_R}  ~~=~~  \frac 1 2\, \wt\p\,.
\end{equation}
	We remind that all derivatives in Eqs.~\eqref{spinor-der} and \eqref{spinor-der-total}
	act to the \emph{left}.

	Quite analogously, we consider the hermitean-conjugate spinors $ \psi_L^\dag $ and $ \psi_R^\dag $,
	for which
\begin{align}
\notag
	\p_{\psi_L^\dag}	& ~~\propto~~	\p_{\xi_L^*}  ~~+~~  \jj\,\p_{\chi_L^*}\,,
	\\[3mm]
\label{diff-psiLR-dag}
	\p_{\psi_R^\dag}	& ~~\propto~~	\lgr \p_{\xi_R^*}  ~~+~~ \jj\,\p_{\chi_R^*} \rgr (-\jj)\,,
\end{align}
	where the derivatives now act \emph{to the right}.
	Notice that the expressions in \eqref{diff-psiLR-dag} are precisely the hermitean conjugates
	of those in Eqs.~\eqref{diff-psiL} and \eqref{diff-psiR}.
	Proceeding the same way as we arrived to Eq.~\eqref{prelim-psi-der} and past it,
	we, now unsurprisingly, find
\begin{align}
\notag
	\p_{\psi_L^\dag}	& ~~=~~		\frac 1 2\, \p^*\, P_L\,,
	\\[4mm]
\label{spinor-der-dag}
	\p_{\psi_R^\dag}	& ~~=~~		\frac 1 2\, \p^*\, P_R\,.
\end{align}
	We stress again that here the derivatives act to the \emph{right}.
	Operator $ \p^* $ is the derivative with respect to the complex conjugates of the
	components \eqref{psiD-as-quat} of $ \psi_D $,
	the latter viewed as a complex quaternion,
\begin{align}
\notag
	\p^*		& ~~=~~	 \p_{t^*}  ~~-~~  \ii\,\p_{x^*}  ~~-~~  \jj\,\p_{y^*}  ~~-~~  \kk\,\p_{z^*}\,,
	\\[3mm]
	\psi_D^*	& ~~=~~  t^*  ~~+~~  \ii\,x^*  ~~+~~  \jj\,y^*  ~~+~~  \kk\,z^*\,.
\end{align}

	Equations \eqref{spinor-der} and \eqref{spinor-der-dag} is our final answer
	for the spinor derivatives.
	The reason is that we already now how to apply a differential operator $ \p $
	or its conjugates in an efficient way --- we have equations \eqref{dq} and \eqref{daq}
	at our disposal.

	If an expression is linear, say in $ \psi_L $, then differentiation
	simply ``erases'' the latter from the expression
	(since the Lagrangian is real, one can always arrange $ \psi_L $ to stand
	on the right of the expression).
	Fermionic expressions are almost always linear in spinors ($ \psi_L^\dag $ or
	$ \psi_L^* $ is an independent variable from $ \psi_L $),
	except for four-fermion interactions.
	Even then, Appendix~\ref{sec-ordinary-quat} explains how to deal with such functions.

	As was shown in Appendix~\ref{sec-ordinary-quat}, an extremum of a real function
	is attained when
\begin{equation}
\label{extremum-eq}
	\p\,f	~~=~~	0\,.
\end{equation}
	Whether the quaternions are ordinary or complex, does not change this.
	As we saw above, such a derivative can be split into left- and right-handed parts,
	each of which provides an independent equation --- for $ \psi_L $ and $ \psi_R $
	correspondingly.
	To see that, it is enough to multiply Eq.~\eqref{extremum-eq}
	on the right by a left- or right-handed projector.
	Thus, we have a recipe for deriving the equations of motion for Weyl spinors
	as well as for Dirac spinors.

	Let us give simple example that we use ---  a Lagrangian of a Dirac fermion,
\begin{equation}
	\cell	~~=~~
			\psi_L^\dag\, i\md\, \psi_L  ~~+~~  \psi_R^\dag\, i\,\wt\md\, \psi_R
			~~-~~  
			\lgr m\, \psi_L^\dag\, \psi_R\, \jj  ~~+~~  \text{q.c.} \rgr  ~~+~~  \text{c.c.}
\end{equation}
	We will need this Lagrangian in the full form,
\begin{align}
\notag
	\cell	&  ~~=~~
			\psi_L^\dag\, i\md\, \psi_L  ~~+~~  \psi_R^\dag\, i\,\wt\md\, \psi_R  ~~+~~
			\wt\psi{}_L\, i\md^*\, \psi_L^*  ~~+~~  \wt\psi{}_R\, i\,\md^\dag\, \psi_R^*  ~~-~~
	\\[2mm]
\label{L-dirac-expanded}
	&  ~~-~~
	m \lgr \psi_L^\dag\, \psi_R\, \jj  ~~+~~  \jj\, \wt\psi{}_R\, \psi_L^* \rgr
	~~+~~
	m \lgr \wt\psi{}_L\, \psi_R^*\, \jj  ~~+~~ \jj\, \psi_R^\dag\, \psi_L \rgr.
\end{align}
	Caution must be executed with respect to signs when conjugating fermions.
	It is advisable that the reader should take a moment to understand why the signs in front of 
	the conjugated terms are as they are shown in Eq.~\eqref{L-dirac-expanded}.

	Let us begin with the kinetic terms first.
	Because of Lorentz transformations (and of gauge transformations, if applicable), 
	$ \psi_L $ and $ \psi_R $ are traditionally written on the right,
	and their hermitean conjugates --- on the left.
	So it is only straightforward for us to apply operator $ \p_{\psi_L}^\dag $
	from \eqref{spinor-der-dag} on the left.
	By construction, operators in \eqref{spinor-der-dag} erase the fermions directly neighbouring them.
	The first kinetic term gives us $ i\md\,\psi_L $.
	What about $ \wt\psi{}_L\, i\md^*\, \psi_L^* $ ?
	Equation \eqref{diff-psiLR-dag} seems to imply that the derivative of $ \psi_L^* $
	with respect to $ \psi_L^\dag $ maybe non-zero.
	While this is actually true, the kinetic term for $ \psi_L^* $ simply gets filtered out by
	the projector $ P_L $ from \eqref{spinor-der-dag}.
	The same happens to the mass term containing $ \psi_L^* $,
	so that only the first mass term contributes, and we arrive to,
\begin{equation}
\label{left-dirac}
	i\md\,\psi_L  ~~-~~  m\,\psi_R\,\jj	~~=~~	0\,.
\end{equation}

	Analogously, the derivative with respect to $ \psi_R^\dag $ of the right-handed kinetic terms
	is $ i\,\wt\md\,\psi_R $.
	As for the mass term bracket containing $ \psi_R^\dag $, one can proceed with it in two ways.
	The first, and perhaps the easiest way, is to re-write it in the form identical to that of the
	left-handed mass term.
	That is, please observe that the last bracket in Eq.~\eqref{L-dirac-expanded}
	is the ``zeroth'' quaternionic component of $ \jj\, \psi_R^\dag\, \psi_L $, 
	and as such, is invariant to cyclic permutations of the latter.
	Using this cyclic symmetry we can pull $ \psi_R^\dag $ to the left,
\begin{equation}
	m \lgr \wt\psi{}_L\, \psi_R^*\, \jj  ~~+~~ \jj\, \psi_R^\dag\, \psi_L \rgr
	~~=~~
	m \lgr \psi_R^\dag\, \psi_L\, \jj  ~~+~~  \jj\, \wt\psi{}_L\, \psi_R^* \rgr.
\end{equation}
	The problem of differentiating with respect to $ \psi_R^\dag $ becomes trivial
	for the first term in the bracket, giving us $ m\, \psi_L\, \jj $.
	The second mass term again gets projected out.
	Altogether, we have
\begin{equation}
\label{right-dirac}
	i\,\wt\md\,\psi_R  ~~+~~  m\, \psi_L\, \jj	~~=~~	0\,.
\end{equation}
	
	The second way of differentiating the mass term with respect to $ \psi_R^\dag $
	is to use Eq.~\eqref{spinor-der-dag} directly.
	This would involve applying derivative $ \p^* $	with the help of identities \eqref{daq}.
	The result will of course be the same.

	We have mentioned here that the derivative $ \p_{\psi_L^\dag}\, \psi_L^* $
	(along with the corresponding right-handed one) is non-zero.
	What is it then?
	Let us see,
\begin{equation}
	\p_{\psi_L^\dag}\, \psi_L^*		~~=~~
	\p_{\psi_L^\dag}\, (\psi_D\,P_L)^*		~~=~~
	\p_{\psi_L^\dag}\, \psi_D^*\,P_R		~~=~~
	\frac 1 2\, \p^*\, P_L\, \psi_D^*\,P_R\,.
\end{equation}
	Using the first identity in \eqref{daq} 
	(for, $ \p^* $ for $ \psi_D^* $ plays the same r\^ole as $ \p $ does for $ q $),
	we find,
\begin{equation}
	\p_{\psi_L^\dag}\, \psi_L^*	~~=~~	-\, \wt P{}_L\, P_R	~~=~~	-P_R\,.
\end{equation}
	It is because of the other factors containing projectors that
	this derivative did not contribute to the equation of motion 
	\eqref{left-dirac}
	(and, correspondingly, the analogous right-handed derivative
	did not contribute to Eq.~\eqref{right-dirac}).


\end{document}